# CompChall: Addressing Password Guessing Attacks

Vipul Goyal[1], Virendra Kumar[2], Mayank Singh[2], Ajith Abraham[3] and Sugata Sanyal[4]

[1] OSP Global, Mumbai, India
vgoyal@ospglobal.com

[2] Crypto Group, Institute of Technology, Banaras Hindu University, India
{virendra.kumar, mayank.singh}@eee06.itbhu.org

[3] School of Computer Science and Engineering, Chung-Ang University, Korea
ajith.abraham@ieee.org

[4] School of Technology & Computer Science, Tata Institute of Fundamental Research, India
sanyal@tifr.res.in

**Abstract**

*Even though passwords are the most convenient means of authentication, they bring along themselves the threat of dictionary attacks. Dictionary attacks may be of two kinds: online and offline. While offline dictionary attacks are possible only if the adversary is able to collect data for a successful protocol execution by eavesdropping on the communication channel and can be successfully countered using public key cryptography, online dictionary attacks can be performed by anyone and there is no satisfactory solution to counter them. This paper presents a new authentication protocol which is called* CompChall *(computational challenge). The proposed protocol uses only one way hash functions as the building blocks and attempts to eliminate online dictionary attacks by implementing a challenge-response system. This challenge-response system is designed in a fashion that it does not pose any difficulty to a genuine user but is time consuming and computationally intensive for an adversary trying to launch a large number of login requests per unit time as in the case of an online dictionary attack. The protocol is stateless and thus less vulnerable to DoS (Denial of Service) attacks.*

## 1. Introduction

Presently, a vast majority of systems use passwords as the means of authentication. Passwords are very convenient for the users, easier to implement and so are very popular also. Although more secure authentication schemes have been suggested in the past, e.g., using smartcards, none of them have been in widespread use in the consumer market. The password based authentication, although very convenient, has some drawbacks due to the very nature of this system. As is obvious, humans have a tendency to choose relatively short and simple passwords that they can remember. Thus the chosen passwords belong to a small domain making them susceptible to exhaustive search or dictionary attacks [2, 3]. There are several instances of such attacks on various systems throughout the world [1].

Password based systems mainly suffer from offline and online dictionary attacks. In an offline dictionary attack the adversary eavesdrops on the communication channel to record data for a successful protocol execution. The adversary then goes offline and tests passwords against the recorded protocol execution data without contacting the server at all. In an online dictionary attack, the adversary tries the possible passwords by attempting to logging in to the server online. Offline dictionary attacks, although severe, can be prevented by various protocols using public key cryptography suggested in the past. The first password based authentication protocol secure against offline dictionary attacks, called EKE, was designed by Bellovin and Merritt [6]. Since then a number of excellent protocols addressing this problem have been proposed [4]. However, no satisfactory measures to curb online dictionary attacks have been suggested so far. There are some methods to deal with them but some of them have security flaws and the others are impractical in terms of usage. Our discussion will be mainly centered on online dictionary attacks and measures to curb it. The proposed protocol employs fast one way hash functions [11] and reduces the number of possible password guesses in a given time period. This is done by asking the client to compute the response for a given challenge. The computation of this response is designed to be a time consuming operation. Special care is taken to ensure that the client is not able to reuse the computation and to make the protocol perfectly stateless.

The rest of the paper is organized as follows:

In Section 2 we discuss the existing protocols, their strengths, weaknesses and flaws (if any). Sections 3 and 4 are dedicated to the proposed protocol where we discuss the basic idea behind the design of the protocol and then discuss the protocol at length. In Section 5 we discuss a few enhancements and modifications for use in specific situations. Finally we conclude the paper in Section 6.

## 2. Related Research

Password based systems are vulnerable to online dictionary attacks. These attacks are difficult to curb and hence pose a major problem in the functioning of password based systems. Countermeasures adopted to prevent the online dictionary attacks are many a times expensive and yet not very effective. Some of the measures adopted to prevent this attack (with their drawbacks) are as follows:

### 2.1. Account Locking

After a few fixed number of unsuccessful login attempts, the account of the particular user is locked for some time. This is certainly helpful in preventing the online dictionary attacks by limiting the number of wrong password guesses in a given time period. However, if adapted, account locking makes the system vulnerable to denial of service (DoS) attacks in which an adversary may launch login requests with random passwords to lock a user's account. Thus, the genuine users are deprived of the service in that period. Yahoo!, for example, reports that users, who compete in auctions, use these methods to block the account of other users who compete in the same auctions. This attack may be worrisome to mission critical applications, for example to enterprises whose employees and customers use the web to login to their accounts. In a similar manner, distributed denial of service (DDoS) attack may also be launched on a system employing the account locking feature. In this, the attacker could plant hidden agents around the web and all the agents would start operating at a specific time. Thus, they would block a large proportion of the accounts of the attacked server by trying to login into accounts in that server using random passwords.

Another major drawback of the "account locking feature" is that since it causes user accounts to be locked, either by mistake (e.g. by users that do not type their passwords correctly) or as a result of dictionary attacks, the service provider must operate customer service centres to handle calls from users whose accounts are locked. The cost of running these centres is high, and is estimated to cost more than $25 per customer call. Imagine that each user locks his account once every five years, then the service cost, per user, per year, is at least $5. A news article [1] suggested that eBay had not implemented account locking features due to the costs of operating customer support centers.

An option here for the service provider could be to automatically unlock the account after a fixed amount of time (e.g. 12 hours). But then, it is easy for anyone to keep the account of a customer always locked (e.g. by using programs which send login requests with random passwords after every 12 hours) and thus totally depriving the customer of the service.

Despite the above serious problem, account locking is still a commonly adopted countermeasure against online dictionary attacks. Several major web based service providers use this approach to counter online dictionary attacks.

### 2.2. Delayed Response

In this scheme, the server provides a delayed response to the user request, say for example, not faster than one answer per second. This may prevent an attacker from checking sufficiently many passwords in a reasonable time.

This scheme is very effective for local machines in which the user has to login to the computer using a physically attached keyboard. However, it is ineffective in a network environment. The attacker can try many login attempts in parallel and circumvent the timing measure using the fact that user logins are typically handled by servers that can handle many login sessions in parallel. For example, the attacker can send a login attempt every 10 milliseconds, thus obtaining a throughput of 100 login attempts per second, regardless of how long the server delays the answer to the login attempt. This scheme also suffers from global password attacks. An attacker may be interested in breaking into any account in the system rather than targeting a specific account. A system that has many user accounts and enables logins over a network accessible to the adversary suffers from such attacks.

## 2.3. Use of CAPTCHA

CAPTCHA (Completely Automated Public Turing Test to tell Computers and Humans Apart) is a scheme [10] which offers a challenge to the user attempting to login. These challenges, for example a distorted and cluttered image of a word with textured background, are easy for humans to respond but rather difficult for computers to answer. It is worthwhile noting here that an online attacker is essentially a programmed computer.

Until recently, this scheme was an effective countermeasure against online dictionary attacks. However, due to recent developments in Artificial Intelligence and Computer Vision, programs are available which can quickly interpret and answer these challenges. EZ-Gimpy and Gimpy for example are word based CAPTCHA's that have been broken by Mori and Malik at UC Berkeley Computer Vision Group [5]. Due to these developments, even CAPTCHA is no longer considered to be a secure technique to prevent online dictionary attacks.

From the above discussion, it is clear that a better and elegant method for solving this pressing problem is required.

## 3. A Brief Idea of the Protocol

The proposed protocol attempts to eliminate the possibility of a large number of password guesses in a small time interval by making guessing a time consuming and costly process. Further, the proposed protocol is stateless and thus less vulnerable to DoS attacks [7]. The protocol does not use public key cryptography. This means that the protocol is vulnerable to offline dictionary attacks [4] if an adversary records data for a successful protocol execution by eavesdropping on the communication channel. In order to resist offline dictionary attacks, the server and client may first establish an SSL connection and the session key could be used to encrypt different messages of the protocol. Major web based service providers like Yahoo! and Hotmail already use SSL for protecting login data in transit. Hence, this does not seem to require any infrastructure changes. In cases where performance degradation due to public key cryptography is a concern, we provide a variant of our protocol which makes it very difficult to launch offline dictionary attack.

The proposed protocol uses only fast one way hash functions [11]. The user and the server are required to perform a few hash computations for each login attempt. The system is deliberately made time consuming and computationally intensive for the client to ensure that it is not able to make a large number of authentication requests per second. However, our system is extremely efficient for the server.

In the following section, we discuss our protocol at length.

## 4. The Protocol Description

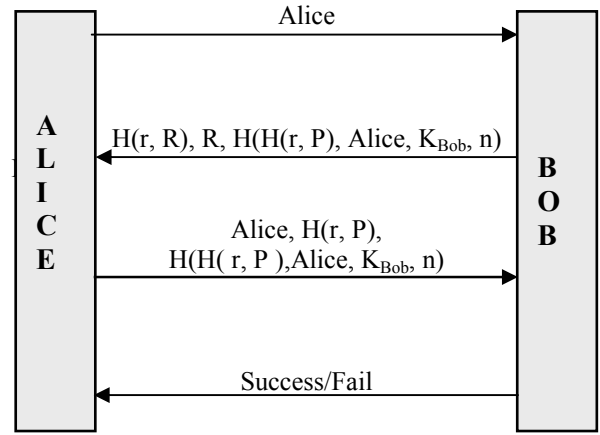

**Figure 1. Illustration of the different passes of the protocol**

The following notations have been used:

- $K_{Bob}$    Secret key of the server Bob, known only to him and no one else
- $P$    Password of the user
- $n$    Number of unsuccessful login attempts to be stored by the server
- $r$    A 20-bit random number
- $R$    A 128-bit random number
- MAC    Message Authentication Code to be sent by the server to the client
- $H(X)$    Hash value of $X$ using a one way collision resistant hash function

This is a four pass protocol with two of the four messages being simple message exchange without any encryption. The rest two involve hash computation once by the user and once by the server. Figure 1 shows the different passes of the protocol. The server presents a challenge to the client with the client login attempt being accepted only if it correctly computes the response to the given challenge. This computation can easily be increased or decreased by the server at will. The proposed protocol is hereby described followed by a brief discussion of the different security measures taken to prevent the major threats. Throughout the discussion, it is assumed that the user is Alice (A) and the server is Bob (B).

### 4.1. Protocol Description

1. A → B:  Alice
   This is a simple login request by the user Alice.

2. B → A:  H(r, R), R, H(H(r, P), Alice, $K_{Bob}$, n)
   In response to the request sent by the user, the server sends a challenge H(*r, R*), the value of R and the message authentication code (MAC) *H(H(r, P)*, Alice, $K_{Bob}$, *n*). The challenge H(*r, R*) is the hash of concatenation of two random numbers *r* (20-bit) and *R* (128-bit). The user is required to compute *r* from the given hash value and the value of R. *r* may be any possible 20-bit number. We will discover shortly the purposes for which *r* and *R* have been used. The third part of the message, MAC is again a hash value and unintelligible to anyone other than the server. This hash can be regenerated only by the server as the secret key $K_{Bob}$ is known only to the server. Note that the client does not use this MAC in anyway. It only has to return the supplied MAC to the server in the next step so that the server does not have to store it. This MAC is used by the server to check the correctness of the value of *r* found by the user and also for the freshness of the message when the user replies with the message 3 as we will see later.
   To find out the value of *r*, the user has to check the hash values of all the possible 20-bit numbers appended with the value of *R*. This process is computationally intensive and may require considerable time (about 5 sec or even more depending on the system used). If instead of two random numbers, only one large random number is used then this computation time is very large and hence the user will be over burdened, which is undesirable. Further, if only a small 20-bit random number is used, then the attacker might store the hash values of all the possible 20-bit random numbers and could easily bypass the computation involved by simply looking for the correct value of *r* from the corresponding stored hash values. The use of two random numbers one of 20 bits and the other of 128 bits thus fulfils two purposes. First, it gives just the right amount of computation to the user so that the online dictionary attacks are effectively countered without inconvenience to a legitimate user. Secondly, it prevents the possibility of pre-computation of hash values of all possible 20 bit numbers. Thus, the number *R* effectively acts as a salt in the computation of the number *r*. The user, after receiving the second message, does the required computation to find the value of *r* after which it proceeds with the third message.

3. A → B:  Alice, H(r, P), MAC
   In order to make the protocol stateless, this step has been made independent of the previous steps, i.e., the client initiates the connection again after doing the required computation and starts with the 3$^{rd}$ step of the protocol directly.
   The user, after receiving the second message, computes the value of *r* from the given values and then sends her identity, hash of the computed *r* concatenated with the password *P*, and the MAC. In the message, *r* and *P* have been hashed instead of sending them directly in plaintext. This is to make the protocol secure against an eavesdropper.
   The server, after receiving this message, finds out the hash of the sent H(*r, P*) appended with the id of Alice, the secret key ($K_{Bob}$) and the stored value of *n*. It then compares the obtained hash value with the sent MAC. If they match, the login attempt is successful, else the login attempt fails and the server increments the value of *n*. The use of the MAC is that it authorizes the supplied r to be the response of a challenge generated by the server and prevents the replay attack in which the attacker may use the same set of values again and again. We have used the value of *n* (number of unsuccessful attempts) in the computation of the MAC. So, a repeated use of message 2 is not possible as *n* increments on every unsuccessful attempt.
   Here, an important point to observe is that *n* does not increment on a successful attempt. This is an interesting feature making the protocol friendly to the legitimate users. This means that if the user was successful in his last login attempt, she would be allowed to bypass the computation involved by reusing the last computation. Thus a legitimate user may actually be required to perform the computation only the first time she tries to login. For every subsequent login attempts, the last computation could be reused as long as the login attempt does not fail.
   By the use of MAC, the server is also relieved from the burden of storing the current value of *r* and *R* for checking the correctness of the value sent by the client. This makes the protocol perfectly stateless.

4. B → A:  Success/Fail
   This is a simple reply by the server indicating whether the information provided by the user was correct or incorrect. If found correct, the login attempt is successful otherwise the user has to start all over again with the first message.

Thus, for every login attempt, the user has to compute the value of r to answer the server's challenge. This computation requires time which may vary from computer to computer. The computation time can be adjusted by simply varying the size of number *r* to keep pace with the computational capacity as it increases with time. This computation time is to discourage the online dictionary attack in which a machine launches thousands of login requests in seconds. By using this technique, the number of authentication requests possible in a given period of time reduces significantly thereby making the process of launching attacks costly and time consuming.

## 4.2. A Brief Security Analysis

In order to better understand the protocol, we discuss the various ways in which an adversary may try to defeat the scheme.

In the proposed scheme, the server is not required to store either *r* or *R*. It verifies the values supplied by the user in message 3 only using the supplied MAC. Thus, an attacker might try to use the same MAC and hence reuse the computation for different login attempts. However, such an attempt is countered by our protocol. The server uses the stored *n* to compute the MAC. Since the value of stored *n* would be more than the value of *n* in the sent MAC, the two MACs would not match. Hence, the attempt to reuse the computation fails. An attacker under no circumstances will be able to change the MAC for different set of values of H(*r, P*) and *n* since the secret key used in the MAC is not known to any entity except the server.

In a similar way, an attempt to use the same value of message 3 (computed for a particular user id) for a different user id will fail since the user id is also used in the MAC computation. Clearly, the only way in which a computation may be reused is to reuse it for the same user id and value of *n*. This means that the computation can be reused in case the last login attempt was successful.

## 5. Enhancements and Modifications

The protocol presented in Section 4 does not take care of the situation in which the server itself may be compromised. This is because the server is required to store the user password in plaintext since it used for the computation of H(*r, P*) in the computation of MAC. The protocol can be augmented so that the server only stores a one way hash H(*P*) of the password and the authenticating user is required to have knowledge of the actual password itself. Straightforward techniques to do this are possible if it is acceptable for the client to send the password in plaintext. However, this would facilitate replay attacks if the protocol execution is not protected by SLL.

Our augmentation is relatively complex but does not make it mandatory to use SSL protection. It employs the concept of lamport hashes [8]. To begin with, the server stores $H^m(P)$ (which is the $m^{th}$ hash of P) and the user is required to supply $H^{(m-1)}(P)$ as a password. Once the user has successfully logged in, the stored $H^m(P)$ is replaced by the supplied $H^{(m-1)}(P)$. Thus, next time the user would be required to supply $H^{(m-2)}(P)$. This process continues for m successful login attempts. Although, it may seem that the user is required to re-initialize the system by choosing a different password after m successful logins, there are efficient recently designed techniques [9] which allow infinite number of login in the lamport system. Messages for the $i^{th}$ execution of the protocol are given below -

A → B: Alice
B → A: H(r, R), R, MAC
A → B: Alice, r, $H^{(i-1)}(P)$, MAC
B → A: Success/Fail
MAC = H(r, $H^i(P)$, Alice, $K_{Bob}$, n)

Thus we have augmented the protocol in such a way that neither the server is required to store plaintext password, nor the password is transmitted in plaintext.

As discussed earlier, it may be desirable to resist offline dictionary attacks without using SSL or other public key cryptographic techniques due to efficiency concern. Although this is not possible theoretically [4], we design a variant which makes it very difficult to launch successful offline dictionary attacks. A minor variation in the messages produces interesting results. In message 2, if H(*r, R*) is changed to H(*r, P, R*) (with other things unchanged), then the protocol is effective in preventing offline dictionary attacks. The protocol execution is as follows -

A → B: Alice
B → A: H(r, P, R), R, MAC
A → B: Alice, H(r, P), MAC
B → A: Success/Fail
MAC = H(H(r, P), Alice, $K_{Bob}$, n)

Now, let's try to analyze the system assuming that the SSL session key has not been used to protect the different passes of our protocol. The only thing unknown to the user is *r*. So the required computation is similar to as in original protocol. Taking the typical amount of time required for a hash computation to be t = 0.005 ms on today's machines, the maximum time required by the user to compute *r* will be $2^{20}*t$ i.e.

$2^{20}*0.005*0.001$ (= 5.24288 seconds) which is 5 seconds approximately.

Let $n$ be the average number of guesses in an offline dictionary attack before the actual password is found out. Now, since the attacker does not know the value of $P$ as well as $r$, he will require $2^{20}*n$ hash computations to find the correct values of $r$ and $P$ from message 2 (i.e. he will have to try all possible combinations of passwords and 20-bit digits). Taking the estimated value of $n$ to be 10 million [12], the time required will be:

$2^{20}*10,000,000*0.005*0.001$ seconds = 52428800 seconds = 1.6625 years.

It is worth noting that in the original protocol proposed in Section 4, the corresponding time to launch successful offline dictionary attack is ($2^{20}*t + n*t$). Evaluating this expression, we get the time to be (5.24288 + 50) = 55.24288 seconds.

Thus, it is clear that this variant is quite effective in the prevention of offline dictionary attacks. This variant may be used when the protocol execution is not protected by SSL due to performance concerns.

## 6. Conclusion

In this paper, we addressed the problem of online dictionary attacks and presented an authentication protocol to counter the same. In the protocol, the client is required to compute the response to the presented challenge. Computing this response is deliberately designed to be a time taking operation thus ensuring that the client is not able to launch a large number of login requests in a small amount of time. The protocol is designed in a fashion such that the computation of this response does not poses any problems for a legitimate user since she may reuse the last computation, but is time consuming and costly for an adversary trying to launch thousands of login requests per second. Finally, we constructed two variants of our protocol. The first one deals with augmenting the protocol so that the server is not required to store the password in plaintext. The second one is concerned with removing offline dictionary attacks in case the public key cryptography protection is not used.

Future work involves modifying the protocol such that the size of $r$ and hence the required computation increases dynamically as the server encounters a large number of unsuccessful attempts in a small amount of time. Finally, the presented technique could be used to address the problem of eliminating more general denial of service attacks on web servers by limiting the number of requests per second in a similar fashion without losing statelessness.